# Opportunities and Challenges in MOCVD of β-Ga2O3 for Power Electronic Devices

M.A. Mastro, J.K. Hite, C.R. Eddy, Jr., M.J. Tadjer
U.S. Naval Research Laboratory, Washington, DC 20375, USA

S.J. Pearton
Department of Materials Science and Engineering, University of Florida, Gainesville, FL 32611, USA

F. Ren
Department of Chemical Engineering, University of Florida, Gainesville, FL 32611, USA

J. Kim
Department of Chemical and Biological Engineering, Korea University, Seoul 02841, Korea

**Abstract**
Recent breakthroughs in bulk crystal growth of β-Ga$_2$O$_3$ by the edge-defined film-fed technique has led to the commercialization of large-area β-Ga$_2$O$_3$ substrates. Standard epitaxy approaches are being utilized to develop various thin-film β-Ga$_2$O$_3$ based devices including lateral transistors. This article will discuss the challenges for metal organic chemical vapor deposition (MOCVD) of β-Ga$_2$O$_3$ and the design criteria for use of this material system in power electronic device structures.

**Properties**

β-Ga$_2$O$_3$-based transistors and diodes possess fundamental electronic properties that make them ideal candidates for high power devices (Table 1). A number of these properties derive directly from the wide band-gap of β-Ga$_2$O$_3$ ($E_g$ = 4.85 eV) including an exceptionally high electric breakdown field (approximately 8 MV/cm). This high breakdown field allows β-Ga$_2$O$_3$-based devices to be biased at a high drain voltage ($V_{break-down}$ >> 10kV) while maintaining a large dynamic range. Furthermore, the wide band-gap of β-Ga$_2$O$_3$ allows device operation at elevated temperature without degradation. Additionally, Ga$_2$O$_3$ has a high saturation electron velocity ($v_{sat}$ = 2 x 10$^7$ cm/s), which is partially accountable for the high current density, $I_{max}$ (where $I_{max}$ ≈ qnv$_{sat}$, q is the elementary charge, and n is the charge density) in devices. [1]

Power semiconductor devices, used in three-terminal switches or two-terminal rectifiers, when forward biased should have minimal resistance in the on-state, $R_{on-sp}$, and support a large blocking voltage, $V_B$, in the off-state. [3] In a standard device design, increasing the thickness, $L_N$, or decreasing the doping, $N_d$, of an n$^-$ drift region increases the on-resistance as described by

$$R_{on-sp} = \frac{L_N}{q\mu_n N_d}$$

Avalanche breakdown occurs when the electric field in the deletion region exceeds the material dependent critical value, $E_c$. [4] For an abrupt junction, the depletion layer extends almost entirely in the lightly doped side as described by

$$W_D = \sqrt{\frac{\varepsilon_S V_B}{q N_d}},$$

where $\varepsilon_s$ is the permittivity. [5] The linearly decreasing field across the depletion layer has a maximum at the junction. For a drift layer thickness sufficient to support this depletion width, $L_N \approx W_D$, the maximum breakdown is set by critical electric field, $V_B = \frac{E_c W_D}{2}$.

| Properties | Si | GaAs | Diamond | 4H-SiC | GaN | $Ga_2O_3$ |
|---|---|---|---|---|---|---|
| Bandgap $E_g$ [eV] | 1.12 | 1.42 | 5.5 | 3.25 | 3.4 | 4.85 |
| Dielectric Constant, ε | 11.8 | 12.9 | 5.7 | 9.7 | 9 | 10 |
| Breakdown Field, $E_c$ [MV/cm] | 0.3 | 0.4 | 20 | 2.5 | 3.3 | 8 |
| Electron Mobility, μ [cm²/V·s] | 1500 | 8500 | 4500 | 1000 | 1250 | 250 |
| Maximum Velocity, $v_s$ [$10^7$ cm/s] | 1 | 1 | 2.5 | 2 | 3 | 2 |
| Thermal Conductivity, λ, [W/cm·K] | 1.5 | 0.5 | 24 | 4.9 | 2.3 | 0.23 |
| *Figure of Merits / relative to Si* | | | | | | |
| Johnson = $E_c^2 \cdot V_s^2 / 4\pi^2$ | 1 | 1.8 | 27777 | 277 | 1089 | 2844 |
| Baliga = $\varepsilon \cdot \mu \cdot E_c^3$ | 1 | 14.7 | 429378 | 317 | 846 | 3214 |
| Combined = $\lambda \cdot \varepsilon \cdot \mu \cdot V_s \cdot E_c^2$ | 1 | 3.7 | 257627 | 248 | 353 | 37 |
| Baliga High Frequency = $\mu \cdot E_c^2$ | 1 | 10.1 | 13333 | 46 | 100 | 142 |
| Keyes = $\lambda \cdot [(c \cdot V_s)/(4\pi \cdot \varepsilon)]^{1/2}$ | 1 | 0.3 | 23.0 | 3.6 | 1.8 | 0.2 |

Table 1. Properties of relevant semiconductor materials and normalized unipolar power-device figures of merit (FOM). The Johnson FOM describes the power-frequency capability, the Baliga FOM gives the specific on-resistance in the drift region, the combined FOM combines the power, frequency, voltage metrics, the Baliga high-frequency FOM provides a measure of switching losses, the Keyes FOM describes the thermal capability to handle high power density at high frequency. The Johnson and Baliga FOMs are remarkably high for $Ga_2O_3$. [2]

Relating these equations gives the maximum blocking voltage inversely related to the doping density in drift layer by

$$V_B = \frac{\varepsilon_S E_c^2}{2qN_d}$$

Again combining these equations shows the inherent tradeoff between on-resistance and blocking voltage

$$R_{on-sp} = \frac{4V_B^2}{\varepsilon_S \mu_n E_c^3},$$

where the denominator of the equation is the Baliga figure of merit. [6] The simplest method to break this design tradeoff is to move to a semiconductor material with a higher critical electric field (Figure 1).

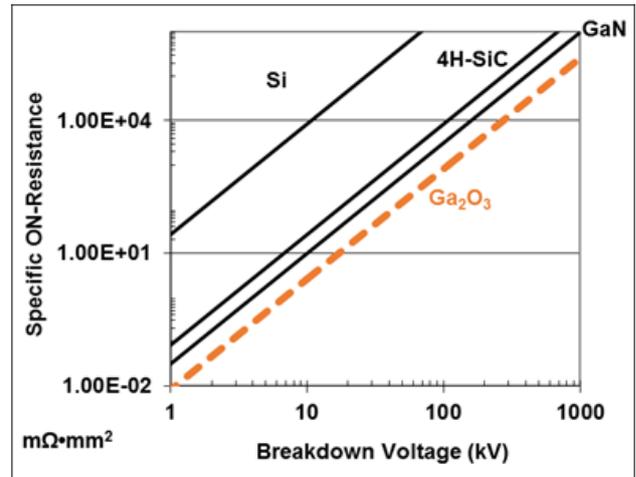

Fig. 1. Each material faces an inherent tradeoff of specific on-resistance to breakdown field as $R_{on-sp} \sim V_B^2$. The high critical electric field of $Ga_2O_3$ provides an inherent advantage compared to GaN and SiC given that $R_{on-sp} \sim E_c^{-3}$. [7]

Examination of the bandgap and breakdown field for various semiconductors reveals a simple relationship given by $\varepsilon_c = a(E_g)^n$ where a and n are fitting parameters. Specifically, the parameters for indirect semiconductors are a = $2.38 \times 10^5$ and n = 1.995, direct semiconductors are a = $1.73 \times 10^5$ and n = 2.506, and all semiconductors are a = $1.75 \times 10^5$ and n = 2.359.

[8] The generally accepted value for breakdown field of β-type of $Ga_2O_3$ is 8 MV/cm although this exceptionally high value has not been experimentally confirmed. [9]

Early demonstrations of high-breakdown β-$Ga_2O_3$ electronic devices are promising, e.g., a critical field strength of 3.8 MV/cm and a 1kV vertical Schottky diode, yet fall short of the predicted levels. Several fundamental material issues are limiting the capability of β-$Ga_2O_3$ for power electronic devices. Experimental measurements of mobility are less than half of the theoretical predictions. Key scattering mechanisms are still unclear including the role of point defects and defect complexes as well as structural stacking faults. Further studies are needed to understand the limitations in saturation electron velocity and breakdown field. [9]

**Crystal Structure of β-$Ga_2O_3$**

Examining the structure of the β-$Ga_2O_3$ crystal helps to frame a number of issues in the growth and behavior of $Ga_2O_3$. The valence band maximum in β-$Ga_2O_3$ forms from weakly interacting O 2p orbital states with contribution of Ga 3d and 4s orbitals while the conduction band minimum forms from Ga 4s states. [10] Closer examination of Figure 2 shows that the $Ga^{3+}$ cations (with a green coloring) have two distinct bonding coordinations. The Ga (I) cation has a distorted tetrahedral coordination with four bonds and the Ga (II) cations has an octahedral coordination with six bonds. Among the common n-type dopants for β-$Ga_2O_3$, Si, and Ge donors prefer the tetrahedral coordination of the Ga(I) site while Sn donors prefers the octahedral coordination of Ga(II) Site

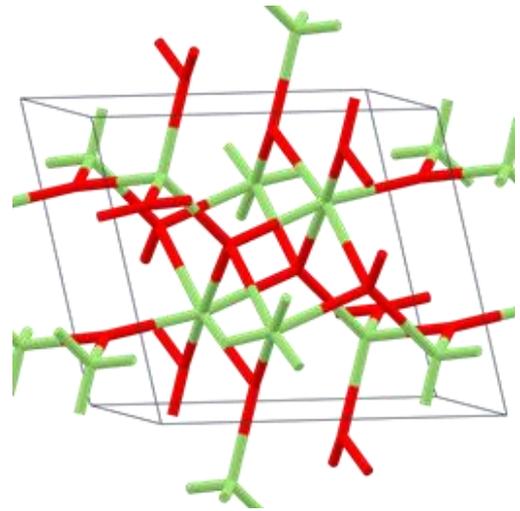

Fig. 2. Bonding structure of monoclinic (C2/M group symmetry) β-$Ga_2O_3$ phase.

Dramatic progress has been made over the past few years in the bulk crystal growth of β-$Ga_2O_3$. The edge-defined film-fed growth (EFG) technique involves pulling the boule along the [010] direction at a rate of approximately 15mm/hr. The EFG process has led to the production of up to 4-inch diameter $Ga_2O_3$ substrates oriented in $(\bar{2}01)$ or (001) planes. The float zone technique is able to produce (100), (010), and (001) oriented boules with a diameter of 1 inch at a growth rate approaching 5mm/hr. The Czochralski technique has produced, at a pull rate of 2 mm/hr, (100) oriented boules of 2-inch diameter with potential for larger diameter boules in the future. [11] The dislocation density of current bulk wafers is of the order $10^3$ $cm^{-2}$, a key result for making large area power devices. [9]

**Homoepitaxy on (100) Plane**

The availability from the Czochralski process of (100) orientated wafers provided a basis for a series of homoepitaxial studies [12, 13] as described in [14]. Schewski et al. found that double positioning (180° rotation) of the monoclinic crystal growth (100) plane leads to twin lamellae formation and stacking mismatch boundaries. This double positioning creates a number of possible monoclinic $Ga_2O_3$ stacking faults including a half unit cell twin layer, a twin layer at surface, and diagonal stacking fault that serves to restore lattice stacking in the direction of growth. [15]

Schewski et al. reported that (100) substrates offcut towards the [00$\bar{1}$] direction provided steps to align the crystal. The growth proceeds in a step-flow manner where impinging adatoms diffuse to a terrace edge. It was observed that an optimal miscut of 6° leads to a density of twins of approximately zero. [15] Conceptually it follows that too small of an offcut can lead to isolated nucleation of islands on large terraces with double positioning and resultant formation of twin lamellae and stacking mismatch boundaries.

In order to describe this process of adatom diffusion to step edges vs. formation of (potentially twinned) islands, Schewski et al. extended a model of Bales and Zangwill [16]. This model required the coupled solution to a set of s ordinary differential equations ODEs for the density of adatoms, <$n_1$>,

$$\frac{1}{F}\frac{d\langle n_1 \rangle}{dt} = \gamma - \frac{D}{F}\xi^{-2}\langle n_1 \rangle - \frac{D}{F}\chi\langle n_1 \rangle - k_1 \langle n_1 \rangle - \sum_{s=1}^{\infty} k_s \langle n_s \rangle,$$

and density of island of size s, <$n_s$>,

$$\frac{1}{F}\frac{d\langle n_s \rangle}{dt} = \frac{D}{F}\sigma_{s-1}\langle n_1\rangle\langle n_{s-1}\rangle - \frac{D}{F}\sigma_s\langle n_1\rangle\langle n_s\rangle + k_{s-1}n_{s-1} - k_s\langle n_s \rangle \quad s = 2,3,...,$$

with equations for flux of incoming adatoms, F, adatom attachment, γ, diffusion to Island, ξ, diffusion to step edge, χ, and deposition on existing island detailed in [15]. The solution to these equations is displayed in Figure 3 at four growth rates that define the adatom flux to the surface.

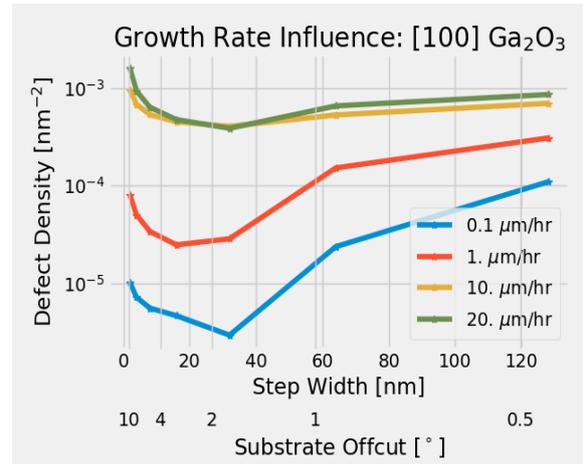

Fig. 3. Calculation of stacking fault density as a function of terrace step width for various growth rates. At low offcut, the adatom surface diffusion length is less than the step terrace width, which leads to island growth with twin and related stacking fault formation. In contrast, a step-bunching mechanism will occur at high offcut. [15]

This calculation provides some guidance on how to control the stacking fault formation for homoepitaxy on (100) plane. Examination of Figure 3 indicates that a layer with minimal stacking faults requires a flux of atoms to the surface in less than the time for these atoms to diffuse to a step edge. The diffusion constant of 7x10$^{-9}$ cm$^2$s$^{-1}$ employed in this calculation is based on a growth temperature of 850°C. Increasing the growth temperature is a clear lever to increase diffusion rate; however, decomposition of the surface is expected at higher temperatures. Nevertheless, reports of Si doping as well as In doping or (In$_x$Ga$_{1-x}$)$_2$O$_3$ alloy formation have shown a surfactant-mechanism that can increase the effective adatom diffusion rate. [17]

**Facet Stability**

A follow-on study by Schewski et al. found that the substrate monoclinic offcut direction of [00$\bar{1}$] and [001] were not equivalent. [18] A Ga$_2$O$_3$ substrate offcut towards [00$\bar{1}$] resulted in steps that reconstruct as the ($\bar{2}$01) facet. Subsequent deposition proceeded in a 2D step-flow manner and the resultant film had high mobility. In contrast, a Ga$_2$O$_3$ substrate offcut towards [001] reconstructed as a twin $(\bar{2}01)$ defect nucleated at a (001)-B step. The film deposited on this substrate possessed a high density of stacking faults (that behaved as acceptor-like electron trap states) and displayed low electron mobility.

An earlier ab-initio studies by Bermudez reported a low surface energy of β-Ga$_2$O$_3$ on the (100) plane. [19] The general understanding of facet energetics was based on this report by Bermudez that only examined that the (100), (010), (001) and (10$\bar{1}$) faces of β-Ga$_2$O$_3$. The energy of these crystal planes is shown in Figure 4 with the inclusion of the $(\bar{2}01)$ plane. [18] The reconstruction of the {001} step edge for the <001> offcut is apparent given that the surface energy of the $(\bar{2}01)$ plane is lower than the (001) plane. As discussed by Schewski et al., the asymmetry of the monoclinic (100) crystal results in an asymmetry of the reconstruction. Specifically the (001)-B step cannot form the $(\bar{2}01)$ plane without a stacking fault. [18]

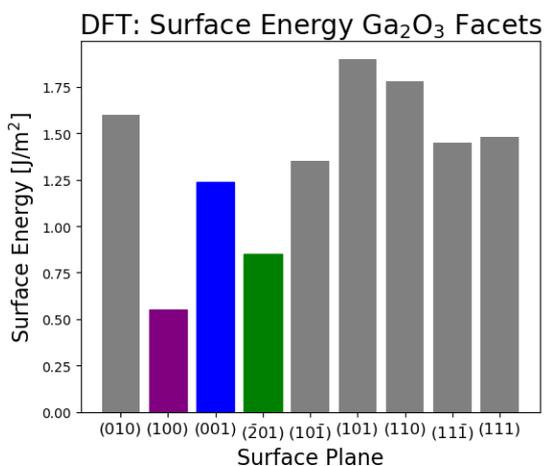

Fig. 4. Density functional theory calculation of surface energy of relevant Ga$_2$O$_3$ surfaces. [18]

Although the preceding discussion was based on homoepitaxy on (100) plane, it is clear that the crystal facet energetics determine the preferred orientations in bulk crystal growth (as discussed above) and the defect formation mechanisms of thin-film growth. Rafique et al. found that hetero-epitaxy on (0001) c-plane sapphire produced $(\bar{2}01)$ oriented Ga$_2$O$_3$ with in-plane rotational domains. The use of (0001) sapphire offcut towards $(11\bar{2}0)$ favored the formation of one domain with the highest mobility found for the film grown on (0001) sapphire with an offcut of 6°. [20] Another effect of the crystal facet energetics was seen in the surface morphology of homoepitaxy (010) Ga$_2$O$_3$, which formed with a striped surface morphology along [001]. [21]

**Carrier Concentration / Compensation**

There has also been progress in the development in the epitaxy of doped β-Ga$_2$O$_3$ by a number of techniques, including MOCVD, HVPE, and MBE with reports of n-type doping over the range 10$^{15}$ to 10$^{19}$ cm$^{-3}$ using Sn or Si shallow donors. [9] It is critical to understand the influence of precursor (trimethylgallium (TMGa) vs. triethylgallium (TEGa)), dopant type (Sn vs. Si) and carrier gas (Ar vs N$_2$). This interplay in the growth environment can be expected given that in β-Ga$_2$O$_3$ the top of the valence and the bottom of the conduction band, respectively, are made up of the anionic (O 2p states with contributions from Ga 3d and 4s orbitals) and cationic states (Ga 4s states). Similarly, the carrier behavior has been shown to dramatically change under annealing, e.g., N$_2$ annealing creates deep acceptor states in n-type β-Ga$_2$O$_3$. [22, 23]

Figure 5 provides a comparison of conductivity for each particular set of precursor and dopant. Baldini et al. found in MOCVD efficient activation of the Si dopant to produce free carriers in the range of 1x10$^{17}$ to 8x10$^{19}$ cm$^{-3}$ in β-Ga$_2$O$_3$ films on (010) β-Ga$_2$O$_3$ substrates. [21] In contrast, incorporation of the Sn dopant was hampered above a concentration of 1x10$^{19}$ cm$^{-3}$. This behavior is framed by the earlier discussion that Si (and Ge) prefer the tetrahedral coordination of Ga(I) site and Sn prefers octahedral coordination of Ga(II) site. Lastly, Baldini et al. reported that a memory effect of Sn in the reactor produced films with an unintentional Sn concentration of approximately 4x10$^{17}$ cm$^{-3}$. [21]

It is commonly understood for MOCVD of compound semiconductor films that the TMGa reaction pathway of Ga(CH$_3$)$_3$ → Ga-CH$_2$(surface) + CH$_4$(g) can leave a high level of carbon in the films. This is in contrast to the TEGa sequential β-elimination pathway of Ga(C$_2$H$_5$)$_3$ → (C$_2$H$_5$)$_2$GaH + C$_2$H$_4$(g) that should yield a semiconductor film with relatively less carbon. Similarly it is generally understood in an MOCVD growth environment with a gallium metalorganic precursor that a decrease in growth temperature will increase the relative level of carbon, which may act as a deep acceptor. This may be especially pertinent given the deposition temperature of β-$Ga_2O_3$ is approximately 300 degrees lower than the typical MOCVD temperature for GaN. Additionally, an increase in growth rate will generally increase the relative level of carbon.

It has been generally observed that MOCVD films produced via a Ga source of TMGa are resistive except at high n-type doping and with $H_2O$ as the oxygen source. Using TMGa and $H_2O$, Gogova et al. doped β-$Ga_2O_3$ with Sn on (0001) sapphire and (100) β-$Ga_2O_3$ substrates (Figure 5). Raman spectroscopy of the films found C-H-related bands. Their analysis stated that Ga vacancy-related defects and the carbon-related complexes act as acceptors compensating for the Sn donors. [24]

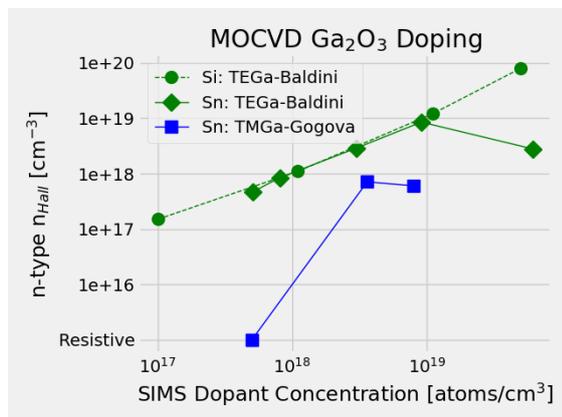

Fig. 5. Comparison of Si and Sn doping during growth of β-$Ga_2O_3$ with TMGa and TEGa precursors (adapted from [21] and [24]). With a TEGa source, moderate levels of Sn and Si dopant demonstrate linear n-type incorporation in β-$Ga_2O_3$.

Tuomisto et al. employed positron annihilation spectroscopy to relate the concentration of negative and neutral vacancies to the conductivity of doped and undoped β-$Ga_2O_3$ thin films. [25] These results as depicted in Figure 6 show that MOCVD with a Ga source of TEGa resulted in a low concentration of gallium vacancies while MOCVD with a TMGa precursor resulted in a high concentration of gallium vacancies. Not shown in the plot is that all films with a vacancy concentration equal to or greater than $1\times10^{17}$ $cm^{-3}$ were insulating. An interesting conclusion from Tuomisto et al. [25] is that growth kinetics and chemical reactions at the MOCVD growth surface dictate the Ga vacancy formation, i.e., not the Fermi level potential of the crystal in a thermodynamic equilibrium condition.

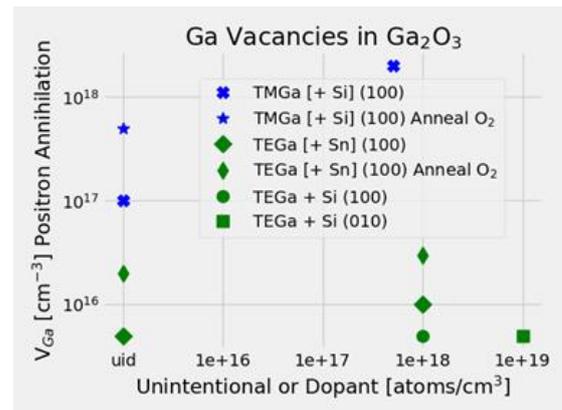

Fig. 6. Comparison of gallium vacancy, $V_{Ga}$, concentration as measured by positron annihilation spectroscopy as a function of dopant concentration for comparing TEGa and TMGa precursors as well as with and without an oxygen anneal (as adapted from [25]). All films deposited with a TMGa precursor displayed a vacancy concentration equal to or greater than $1\times10^{17}$ $cm^{-3}$ and were electrically insulating (not shown).

An underlying mechanism for unintentional conductivity in β-$Ga_2O_3$ was suggested to be based on hydrogen. [26, 27] Interstitial hydrogen ($H_i$) or hydrogen trapped at oxygen vacancies ($H_O$) is predicted to act as a shallow donors. [9] Qin et al. employed infrared spectroscopy to study β-$Ga_2O_3$ annealed in a $H_2$ or $D_2$ ambient. This study found a hidden reservoir of hydrogen that is composed of various hydrogen centers coupled to a gallium vacancy as well as a variety of other species. Vibrational spectroscopy assigned the dominant hydrogen center to a neutral complex composed of two equivalent hydrogen at a relaxed gallium vacancy, $V_{Ga(I)}$-2H. [28]

**Mobility Limits**

The reported theoretical room temperature mobility of β-Ga$_2$O$_3$ is in the range of 200 to 300 cm$^2$/V·s; however, typical experimental mobilities are reported in 50 to 150 cm$^2$/V·s range. [9] Referring to Table 1, the mobility of β-Ga$_2$O$_3$ is low relative to other wide bandgap semiconductors. It is constructive to understand what limits the electron mobility in the β-Ga$_2$O$_3$ crystal. As discussed by Ma et al., β-Ga$_2$O$_3$ has an effective mass, $m_c^*$, of approximately 0.25m$_0$, which is comparable to the effective mass of GaN of 0.21m$_0$ although the mobility of GaN is approximately 1200 cm$^2$/V·s. [29]

It is known that the bonding of Ga and O has a large difference in electronegativity, X$_{AB}$, or similarly a large Pauling ionicity given by $f_P = 1 - exp(\frac{-X_{AB}^2}{4})$. A useful value for predicting the influence of electron to polar optical phonon interaction in polar semiconductors is the Fröhlich coupling constant,

$$\alpha_F = \frac{q^2}{8\pi\varepsilon_0}\sqrt{\frac{2m_c^*}{\hbar\omega_0}}\left(\frac{1}{\varepsilon_\infty} - \frac{1}{\varepsilon_s}\right),$$

where $\hbar\omega_0$ is the polar optical phonon energy, $\varepsilon_\infty$ and $\varepsilon_s$ are the high- and low-frequency relative dielectric constants, and $\varepsilon_0$ is the vacuum permitvitty. The Fröhlich coupling constant is quite large in β-Ga$_2$O$_3$ crystal as can be seen in Figure 7.

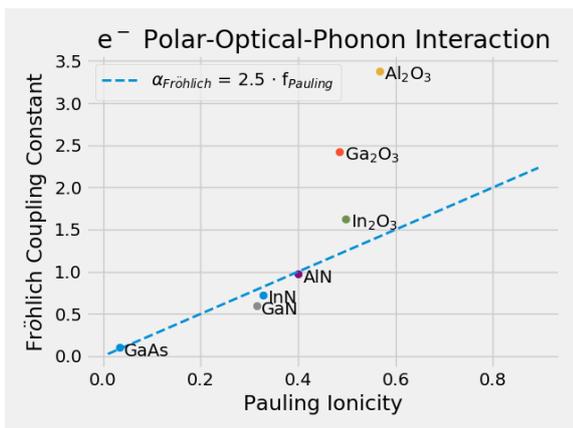

Fig. 7. The exceptionally strong electron to polar optical phonon interaction as measured by the dimensionless Fröhlich coupling constant in β-Ga$_2$O$_3$ greatly exceeds what would be expected from a naive linear prediction given the strong ionic character (as measured by the Pauling ionicity) in the Ga-O bond. [29]

The mobility in semiconductors is controlled by several scattering mechanisms with one mechanism usually dominant for a given set of dopant levels and temperature. It is common in polar semiconductors at room temperature for phonon scattering to limit the mobility at low donor levels while impurity scattering limits the mobility at high impurity levels. The impact of the predicted large Fröhlich coupling constant is apparent in Figure 8a, which displays the mobility of β-Ga$_2$O$_3$ as a function of doping. It is clear in Figure 8a that the electron scattering by polar optical phonons limits the mobility to approximately 200 cm$^2$/V·s for β-Ga$_2$O$_3$ with donor densities less approximately 5x10$^{18}$ cm$^{-3}$.

Examining the mobility at a low donor level in Figure 8b reveals that electron mobility is controlled by polar optical phonon scattering at temperatures above 200K. The general understanding of semiconductor transport is that phonon scattering is inherent to the particular crystal and only marginal improvement may be possible through strain engineering. [30]

In contrast, the mobility at a donor level of 1x10$^{20}$ cm$^{-3}$ is depicted in Figure 8c. As mentioned above, neutral and ionized impurity scattering dominates at above approximately 5x10$^{18}$ cm$^{-3}$. In Figure 8c, the high donor level leaves a large density of neutral impurities that are the dominant scattering source in the temperature range depicted.

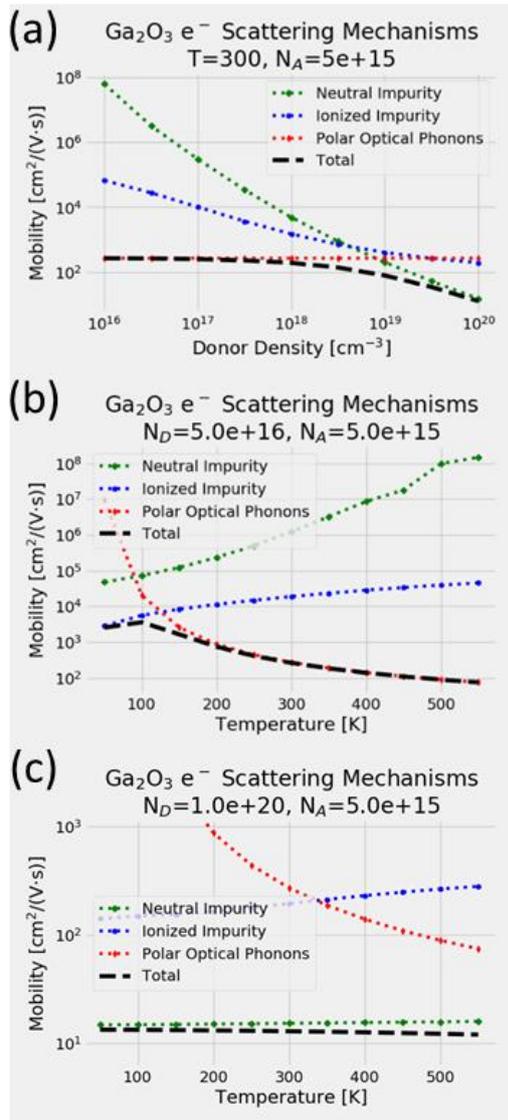

Fig. 8. Calculated mobility and relevant scattering mechanisms in β-Ga$_2$O$_3$ (a) at 300K as a function of donor density, (b) as a function of temperature at a low donor level (5x10$^{16}$ cm$^{-3}$) where polar optical phonon scattering is the dominant mechanism limiting the mobility above 200K, and (c) as a function of temperature at a high donor level (1x10$^{20}$ cm$^{-3}$) where impurity scattering controls the mobility.

The strong influence of neutral and ionized impurity scattering at high doping levels motivated the development of modulation doped heterostructures where the donor source and the transport channel are physically displaced. This design is the basis of the well-known modulation doped field effect transistor (MODFET). This structure is depicted in Figure 9a, where donors in the delta doped portion of the (Al$_x$Ga$_{1-x}$)$_2$O$_3$ barrier are physically displaced from the conductive channel formed in a triangular potential well at the (Al$_x$Ga$_{1-x}$)$_2$O$_3$ / Ga$_2$O$_3$ interface.

A key design criteria in MODFETs is the physical spacing between the two-dimensional electron gas in the channel and the delta doped layer. As is visible in Figure 9a, the electron wavefunction for the conductive channels extends to the donors in the delta-doped barrier. This provides a mechanism for screening of the electrons in the channel by neutral and ionized donors in the barrier. This confinement of carriers is further improved by increasing the aluminum composition in the barrier to increase the conduction band offset between (Al$_x$Ga$_{1-x}$)$_2$O$_3$ and Ga$_2$O$_3$. An additional bandgap engineering constraint is to prevent the formation of a second conductive channel at the delta-doped layer in the barrier.

As discussed earlier, polar optical phonon scattering creates a fundamental limitation that is difficult to mitigate in the mobility of polar semiconductors. Ghosh and Singisetti recently made a dramatic prediction of a screening mechanism for polar optical phonons in Ga$_2$O$_3$. [32] It is known that a 2D electron gas in a semiconductor behaves at a plasmon wave with a characteristic energy defined by the density of electrons. [31] Ghosh and Singisetti studied how this plasmon wave couples to the longitudinal optical phonon modes in Ga$_2$O$_3$. This remarkable study found that at moderate electron densities the plasmon wave will anti-screen the longitudinal optical phonons, which increases the scattering rate; while at high electron densities the plasmon wave will screen the longitudinal optical phonons, which decreases the scattering rate. [32]

This anti-screening / screening behavior is observable in the dot-dashed line in the mobility vs. channel density plot in Figure 9b. Dramatic improvements in mobility are predicted at electron densities above 5x10$^{18}$ cm$^{-3}$. Unfortunately, uniformly doping a single epilayer to achieve these carrier densities will create a high density of impurity scattering sites that will severely limit the mobility as was discussed for Figure 8. Again the solution is to separate the dopants from 2D electrons. Still, the electrons in the triangular potential well can experience scattering by the donors in the delta-doped layer. The mobility accounting for the anti-screening / screening behavior as well as the scattering by the remote impurities is displayed by the dashed line in Figure 9b. As can be seen, there is a clear design tradeoff to minimize the impurity scattering sites vs. increasing the channel density to increase the polar optical phonon screening - as well as standard MODFET design rules. [4, 33, 34]

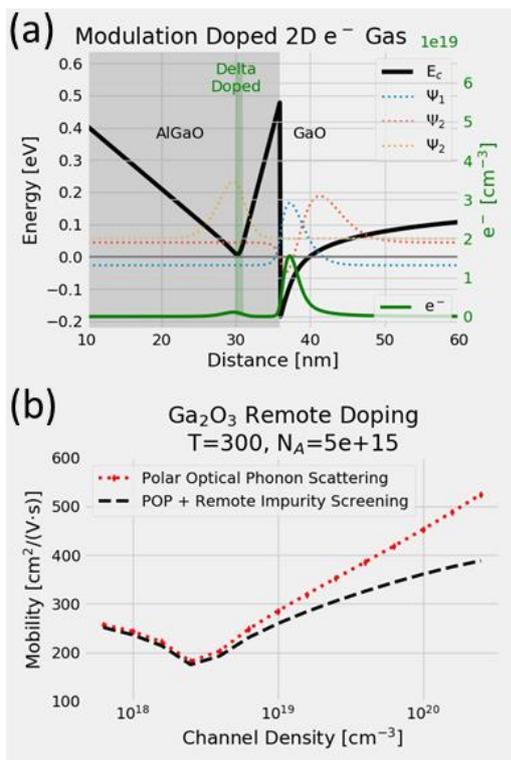

Fig. 9. β-$(Al_xGa_{1-x})_2O_3$ / $Ga_2O_3$ modulation doped structure with a conductive channel formed in a triangular potential well. (a) Calculation of conduction band energy, electron concentration, and electron wavefunction for a structure composed of 35nm delta-doped $(Al_xGa_{1x})_2O_3$ on $Ga_2O_3$. A small portion of the electron wavefunction in the channel extends to the delta-doped donors in the $(Al_xGa_{1-x})_2O_3$ layer. (b) The dot-dashed line depicts a mobility model that accounts for polar optical phonon anti-screening at moderate electron densities and screening at high electron densities. [32] The dashed line depicts the mobility including screening by the remote impurities in the delta-doped layer.

**Alloy Formation**

From the previous section it is clear that research is needed in epitaxy of $(Al_xGa_{1-x})_2O_3$ / $Ga_2O_3$ heterostructures. The $(Al_xGa_{1-x})_2O_3$ alloy is challenging as α-$Al_2O_3$ is stable as the corundum phase (Figure 10) while β-$Ga_2O_3$ is stable as the monoclinic phase.

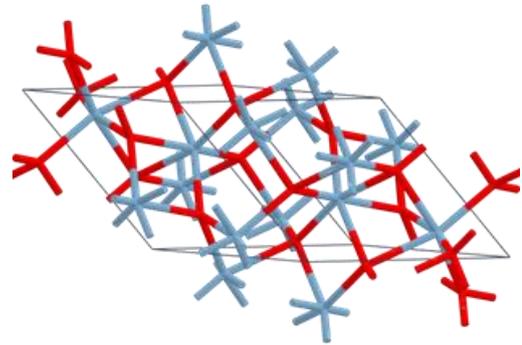

Fig. 10. Bonding structure of the stable corundum α-$Al_2O_3$ (R3c group symmetry) phase where the Al cations are octahedral coordinated with six bonds.

The phase of $(Al_xGa_{1-x})_2O_3$ that forms not only has a different bandgap energy but presents a different transition state. It is known that the β-$Ga_2O_3$ is theoretically an indirect semiconductor but this direct gap transition is so similar in energy that β-$Ga_2O_3$ effectively behaves as a direct gap semiconductor. Examination of Figure 11, which is based on the model in Peelears, et al. [35], shows that the energy difference of the indirect transition compared to the direct transition increases to a significant level with increasing alloy composition. The influence of this effect on device properties is not well studied. Similarly, the separation of the indirect and direct transition energies in the $(Al_xGa_{1-x})_2O_3$ corundum phase is also shown in Figure 11.

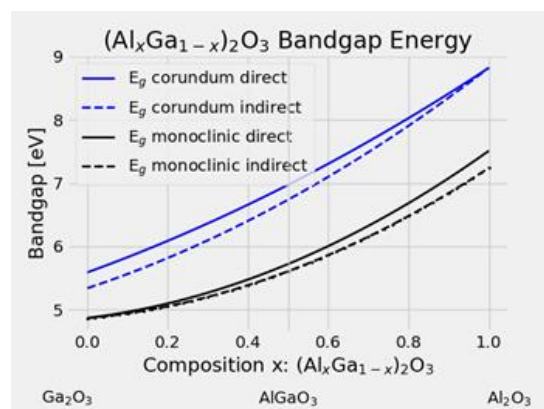

Fig. 11. Direct and indirect bandgap energies of the monoclinic and corundum phases of the $(Al_xGa_{1-x})_2O_3$ alloy. [35]

It is known that forming alloys of $(Al_xGa_{1-x})_2O_3$ is difficult. Examination of Figure 12 based on the enthalpy of formation model of Peelears, et al. [35] provides some insight. The calculation suggests that the monoclinic phase is stable for x < 0.71. The stability local minimum in the formation energy for AlGaO$_3$ (x = 0.5) is logical given the crystal structure of the binary compounds. The bonding structure of the corundum α-Al$_2$O$_3$ phase is composed of Al cations only at octahedral coordinated sites with six bonds each (Figure 10). In contrast, Figure 2 shows that the monoclinic β-Ga$_2$O$_3$ phase is composed of equal parts of Ga (I) cations in a distorted tetrahedral coordination with four bonds and Ga (II) cations in an octahedral coordination with six bonds. At the stability local minima at x = 0.5, the Ga (I) cations are at the tetrahedral coordinate site while, critically, the Al cations are only present at the octahedral coordination site.

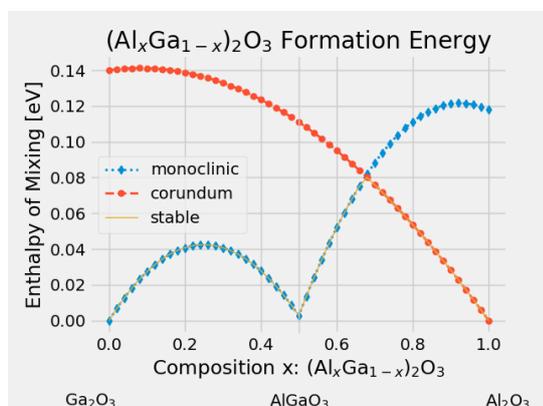

Fig. 12. Theoretical formation energy of $(Al_xGa_{1-x})_2O_3$ based on the model of Peelaers, et al.[35]. The monoclinic phase is stable for x < 0.71 and the corundum phase is stable at higher mole fractions.

Early work by Hill et al. into the equilibrium diagram of Al$_2$O$_3$ in β-Ga$_2$O$_3$ found the presence of a stable phase of AlGaO$_3$. [36] This work reported that this phase required a temperature of 800 °C, which may preclude lower temperature growth techniques such as MBE. The exploration of AlGaO$_3$ growth conditions and the role of strain is an important area of future research.

The addition of $(In_xGa_{1-x})_2O_3$ to heterostructures would also be beneficial for the development of electronic devices. Again, the In$_2$O$_3$ crystal possesses a different stable phase (as seen in Figure 13), which contributes to the difficulty in forming a $(In_xGa_{1-x})_2O_3$ stable alloy. Single crystal monoclinic structures were only reported at low indium content (x < 0.15). At high indium content (x > 0.8) the cubic bixbyite phase is formed while at intermediate values an additional rhombohedral InGaO$_3$(II) crystallographic phase formed. [37] Regardless, as discussed above, reports indicate that indium behaves as a surfactant so its key role may be in improving the diffusivity of the gallium atom during growth.

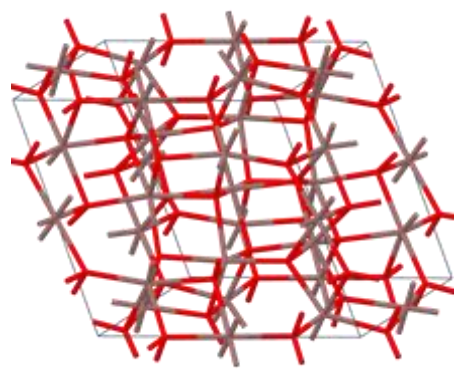

Fig. 13. Bonding structure of the stable cubic bixbyite In$_2$O$_3$ (Ia=3 [206]) group symmetry phase with the In cations in an octahedral coordination with six bonds.

**Conclusion**

Development of β-Ga$_2$O$_3$-based MOCVD technology will enable power electronic devices not possible with other semiconductor materials. The energetics of the crystal facets has a strong influence on growth morphology and defect formation for a given substrate orientation and offcut, growth temperature, and growth rate. It is clear that the growth environment, particularly metalorganic precursor selection and the presence of hydrogen, has a strong impact on the formation of gallium vacancies and the resulting compensation of intentional dopants. These compensating centers may be native defects or complexes from either sublattice. The understanding of the role of hydrogen including in the formation the vacancy complex is still evolving. The mobility in β-Ga$_2$O$_3$ is limited at low donor densities by polar optical phonons. The possibility to screen the longitudinal optical phonon modes by the electron gas plasmon further motivates development of modulation doped (Al$_x$Ga$_{1-x}$)$_2$O$_3$ / β-Ga$_2$O$_3$ heterostructures. To achieve efficient devices based on this structure requires additional development into the growth of (Al$_x$Ga$_{1-x}$)$_2$O$_3$ particularly at high alloy mole fraction.

**Acknowledgments**

The work at NRL was partially supported by DTRA Grant No. HDTRA1-17-1-0011 (Jacob Calkins, monitor) and the Office of Naval Research. The work at UF is partially supported by HDTRA1-17-1-0011. The project or effort depicted is sponsored by the Department of the Defense, Defense Threat Reduction Agency. The content of the information does not necessarily reflect the position or the policy of the federal government, and no official endorsement should be inferred. The work at Korea University was supported by the Korea Institute of Energy Technology Evaluation and Planning (KETEP) and the Ministry of Trade, Industry and Energy (MOTIE) of Korea (Grant No. 20172010104830) and Space Core Technology Development Program (2017M1A3A3A02015033) through the National Research Foundation of Korea funded by the Ministry of Science, ICT and Future Planning of Korea.